# High Phonon Scattering Rates Suppress Thermal Conductivity in Hyperstoichiometric Uranium Dioxide

Hao Ma,[1*] Matthew S. Bryan,[1] Judy W. L. Pang,[1] Douglas L. Abernathy,[2] Daniel J. Antonio,[3] Krzysztof Gofryk,[3] and Michael E. Manley[1*]

[1]*Materials Science and Technology Division, Oak Ridge National Laboratory, Oak Ridge, Tennessee 37831, USA*
[2]*Neutron Scattering Division, Oak Ridge National Laboratory, Oak Ridge, Tennessee 37831, USA*
[3]*Idaho National Laboratory, Idaho Falls, Idaho 83415, USA*

**Abstract**
Uranium dioxide ($UO_2$), one of the most important nuclear fuels, can accumulate excess oxygen atoms as interstitial defects, which significantly impacts thermal properties. In this study, thermal conductivities and inelastic neutron scattering measurements on $UO_2$ and $UO_{2+x}$ (x=0.3, 0.4, 0.8, 0.11) were performed at low temperatures (2-300 K). The thermal conductivity of $UO_{2+x}$ is significantly suppressed compared to $UO_2$ except near the Néel temperature $T_N$= 30.8 K, where it is independent of *x*. Phonon measurements demonstrate that the heat capacities and phonon group velocities of $UO_2$ and $UO_{2+x}$ are similar and that the suppressed thermal conductivity in $UO_{2+x}$ results from high phonon scattering rates. These new insights advance our fundamental understanding of thermal transport properties in advanced nuclear fuels.

* Corresponding author. Email: mah1@ornl.edu
* Corresponding author. Email: manleyme@ornl.gov

## 1. Introduction

Uranium dioxide is the most widely used nuclear fuel and its thermal properties are important to reactor safety and performance[1]. $UO_2$ is able to accommodate a variable stoichiometry, depending on temperature and oxygen pressure[2, 3]. For example, $UO_2$ can be oxidized by water vapor in the rare event of a cladding breach in a reactor, which consequently raises ratio of oxygen to uranium (O/U=2+$x$)[2]. The excess oxygen atoms in hyperstoichiometric uranium dioxide ($x >$ 0) were identified as interstitial defects that form partially ordered clusters[4-6]. In addition, the O/U ratio is one of the most important parameters governing fuel safety because it significantly affects the thermal properties such as the melting point[7] and thermal conductivity[8-12]. For example, Manara *et al.*[7] found the melting point of $UO_{2+x}$ ($x = 0-0.21$) decreases as the O/U ratio increases. White *et al.*[11] reported that $UO_{2+x}$ pellets exhibit a decrease in thermal conductivity with increasing temperature and follow 1/T dependence due to the anharmonic Umklapp phonon-phonon scattering at 363 -1673 K. The thermal conductivity of $UO_{2+x}$ also decreases as $x$ increases, indicating the importance of phonon-defects scattering.

Although there are many studies of the thermal properties of $UO_{2+x}$ at high temperatures (300-1700 K)[7-13], there is little information on the thermal conductivity and phonon properties of $UO_{2+x}$ at low temperatures (below 300 K). The effects of excess oxygen atoms ($x$) on the thermal conductivity of $UO_{2+x}$ at low temperatures, especially around Néel temperature $T_N$= 30.8 K, are of critical importance in bench marking theoretical models. In this study, the thermal conductivity of $UO_2$ and $UO_{2+x}$ single crystals are reported in the temperature range of 2-300 K. We find that $UO_{2+x}$ has a much smaller thermal conductivity than $UO_2$ at all temperatures except near the Néel temperature, where it is unaffected. The phonon density of states (PDOS) and scattering function measured by inelastic neutron scattering (INS) demonstrate that the suppressed thermal conductivity in $UO_{2+x}$ originates with an increased phonon scattering rate with the addition of oxygen interstitial defects.

## 2. Method
2.1 Samples preparation and thermal conductivity measurements

Thermal conductivity measurements were performed in a DynaCool-9 Quantum Design Measurement System. The measurements were performed in the continuous heating mode of the TTO option using a pulse-power steady-state method. The crystals were measured from 2 to 300 K at 0.25 K/min. The oxygen stoichiometry of the $UO_2$ single crystals was set in a thermogravimetric analyzer by controlling the oxygen activity and by adjusting the partial pressure of oxygen at 1273 K. Then, the final stoichiometry of the $UO_{2+x}$ crystals was calculated from the sample weight change relative to the stoichiometric $UO_2$ reference data. The size of these crystals is around 1x1x3 $mm^3$.

2.2. INS measurements

The single crystals used in the thermal conductivity measurements are too small for direct INS measurement on phonon dispersion and especially phonon linewidths. Hence, to characterize the phonon properties, instead INS measurements on powders of $UO_2$ and $UO_{2.08}$ at 77 K and 295 K were performed using the wide Angular Range Chopper Spectrometer (ARCS) at the Spallation

Neutron Source (SNS) of Oak Ridge National Laboratory[14]. The setup of the spectrometer was identical to previously reported PDOS measurements for $UO_2$[15, 16]. An incident neutron energy of $E_i$=120 meV was used, which is high enough to capture the phonon cutoff around 80 meV and allows for summing over enough zones in momentum space to obtain a PDOS. Scattering introduced by the sample can and the cryostat were corrected for by subtracting the corresponding spectra from a duplicate, empty sample can measurement. The corrected scattered neutron intensities, $I(\Phi,t)$, were converted to the scattering function $S(Q,E)$, where Q is momentum transfer magnitude and E is the energy transfer. Other experimental details can be found elsewhere[15, 16] and are not repeated here. The neutron weighted PDOS $g^{NW}(E)$ was then obtained by integrating S(Q, E) over Q values ranging from 4 to 10 Å$^{-1}$ for both $UO_2$ and $UO_{2.08}$, and then correcting for multiphonon scattering (using iterative procedure), the Debye-Waller factor, thermal population factor, and subtracting the elastic peak.

The measured neutron weighted PDOS can be expressed as

$$g^{NW}(E) = \frac{\sigma_U}{M_U} g_U(E) + 2\frac{\sigma_O}{M_O} g_O(E) \tag{1}$$

where $g_i$, $M_i$ and $\sigma_i$ are the partial PDOS, the atomic mass, and the corresponding neutron scattering cross section of element i (i = U or O).[15, 17] In order to extract the neutron unweighted PDOS, we take advantage of the fact that nearly all phonon modes below 25 meV are from uranium and nearly all phonon modes above 25 meV are from oxygen[15, 18, 19].

The specific heat at constant pressure, $C_P$, can be written

$$C_P - C_V = 9\alpha^2 BvT \tag{2}$$

where $C_V$ is the specific heat at constant volume, $\alpha$ is the linear thermal expansion, $B$ is bulk modulus, $v$ is molar volume, and $T$ is temperature[20]. The phonon contribution to the specific heat at constant volume $C_V^{ph}$ can be obtained directly from the neutron unweighted PDOS using:

$$C_V^{ph}(T) = 3k_B \int_0^\infty g_{T_0} \frac{E^2}{(k_B T)^2} \frac{e^{\frac{E}{k_B T}}}{(e^{\frac{E}{k_B T}}-1)^2} dE \tag{3}$$

where $k_B$ and $g_{T_0}(E)$ are the Boltzmann constant and neutron unweighted PDOS at temperature $T_0$, respectively[16].

## 3. Results and discussion

The measured thermal conductivity of $UO_{2+x}$ single crystals as a function of temperature is shown in Fig. 1a. The thermal conductivity of $UO_{2+x}$ for all values of $x$ measured exhibits a double-peak behavior where maxima occur at ~10 and 220 K and a minimum occurs at the Néel temperature $T_N$= 30.8 K, consistent with previous studies on $UO_2$[21-23]. In addition, other than in the vicinity of $T_N$, the thermal conductivity of $UO_{2+x}$ is significantly suppressed compared to $UO_2$ and decreases with increasing $x$ (see Fig. 1b). The decreasing thermal conductivity with increasing $x$ is likely driven by phonon-defects scattering and is similar to the observed high temperature behavior[8-12]. More specifically, the thermal conductivity of $UO_{2.08}$ (2.68 W/(mK)) is 67% the value for $UO_2$ at 77 K (4.46 W/(mK)). Interestingly, the thermal conductivity of $UO_{2+x}$ (~1 W/(mK)) is essentially independent of the $x$ at $T_N$ = 30.8 K, indicating negligible phonon-defects scattering likely due to strong magnetic fluctuations, see Fig. 1b.

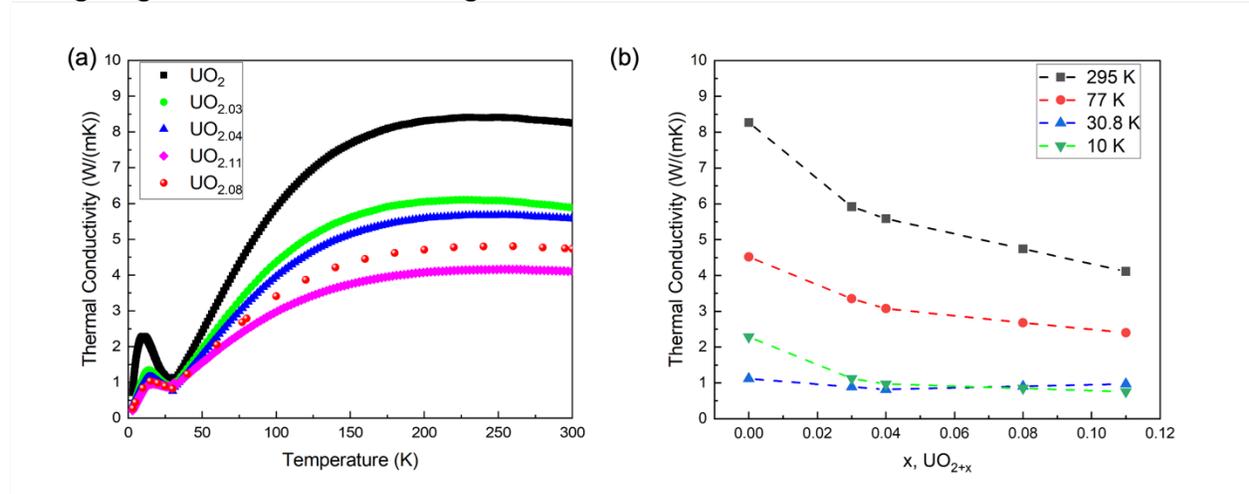

Fig. 1. (a) Measured thermal conductivity of $UO_2$ and $UO_{2+x}$ single crystals as a function of temperature. Thermal conductivity of $UO_{2.08}$ is obtained by interpolation between $UO_{2.04}$ and $UO_{2.11}$ (b) Measured thermal conductivity of $UO_{2+x}$ single crystals as a function of the excess oxygen atoms $x$ at several different temperatures.

To gain deeper insights on the suppressed thermal conductivity in $UO_{2+x}$, we first checked the structure of $UO_{2.08}$ and the measured neutron diffraction pattern of $UO_{2.08}$ in Fig. S1 of Supplementary Materials (SM) shows that $UO_{2.08}$ adopts a superlattice structure which is a mix of $UO_2$ and $U_4O_9$, consistent with reported phase diagram at low temperatures[13, 24, 25]. We then measured their neutron weighted PDOS of $UO_2$ and $UO_{2.08}$ at 77 K and 295 K, as shown in Fig. 2a and 2b. Well-defined zone-boundary phonon peaks are observed at energies of 12, 21, 33, 56, and 72 meV for both $UO_2$ and $UO_{2.08}$ at 77 K and 295 K. With reference to the phonon dispersion of $UO_2$[15, 17], the uranium-dominated transverse acoustic (TA) and the longitudinal acoustic (LA) zone-boundary phonon energies correspond to the 12 and 21 meV peaks. These peaks show trivial difference between $UO_2$ and $UO_{2.08}$. The oxygen-dominated transverse optical (TO1 and TO2) and the longitudinal optical LO2 zone-boundary phonon energies correspond to the remaining 33, 56, and 72 meV peaks, respectively. In contrast, these peaks in $UO_{2.08}$ are less intense and slightly broader than that in $UO_2$. In addition, there are some nonnegligible intensities above phonon cutoff (78meV), which is consistent with nonlinear propagating modes (NPM) identified in a previous study[26].

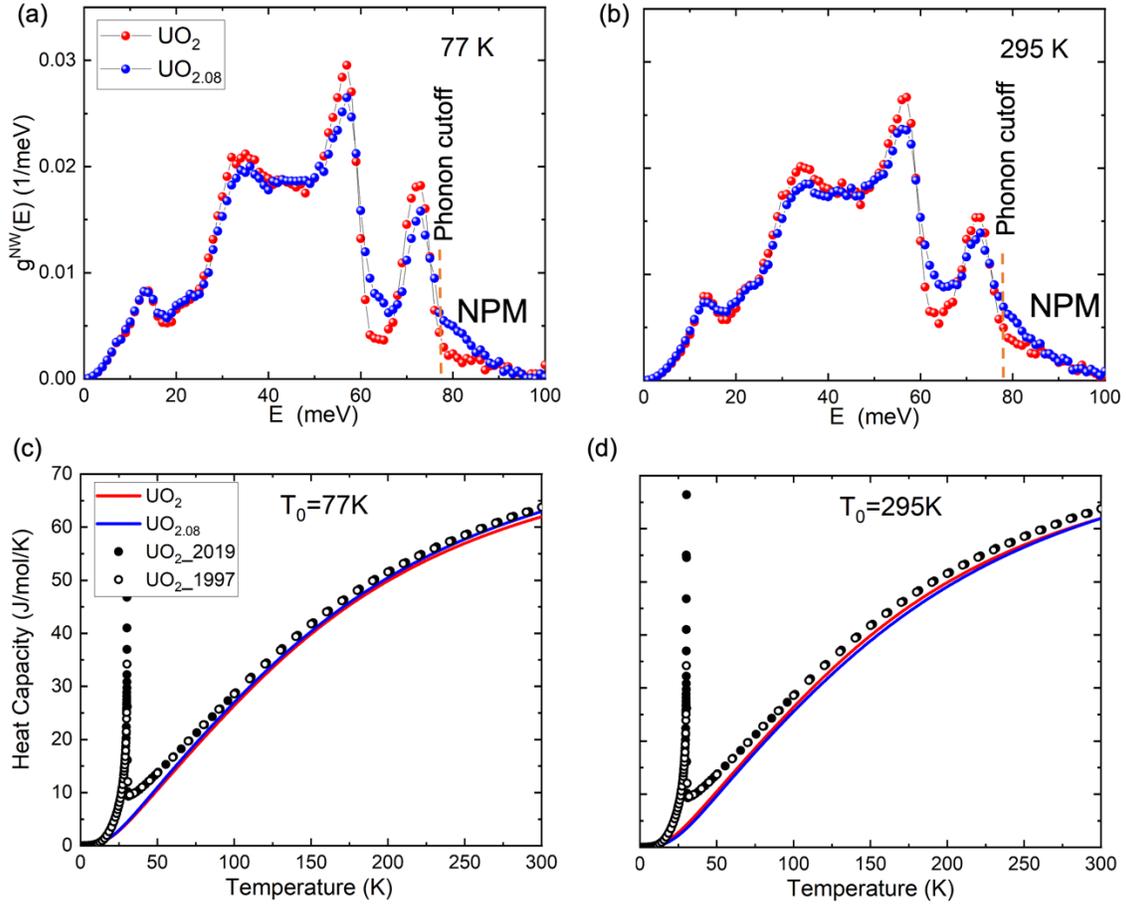

Fig. 2. The neutron weighted PDOS of $UO_2$ (red spheres) and $UO_{2.08}$ (blue spheres) powders measured by INS at (a) 77 K (b) 295 K; The specific heat capacity ($C_P$) of $UO_2$ (red line) and $UO_{2.08}$ (blue line) with respect to temperature were calculated using unweighted PDOS at (c) $T_0$ =77 K and (d) $T_0$ =295 K. Solid and open black circles are previous experimental $C_P$ of $UO_2$ single crystal[16] and sintered powder[27]. A sharp spike at $T_N$ = 30.8 K is a result of the Néel transition from an antiferromagnetic state to a paramagnetic state.

The calculated specific heat capacities ($C_P$) of $UO_2$ and $UO_{2.08}$ using the neutron unweighted 77K and 295K- PDOS based on equation (2) and (3) are shown in Fig. 2c and 2d. Our calculated $C_P$ of $UO_2$ shows good agreement with the experimental $C_P$[16, 27], except in the vicinity of $T_N$ where magnetic contributions are large[28-30]. Moreover, $C_P$ of $UO_2$ and $UO_{2.08}$ are almost same (less than 2% difference) over the entire temperature range of 2-300 K. Note that their $C_V^{ph}$ also show trivial difference as shown in Fig. S2 of SM.

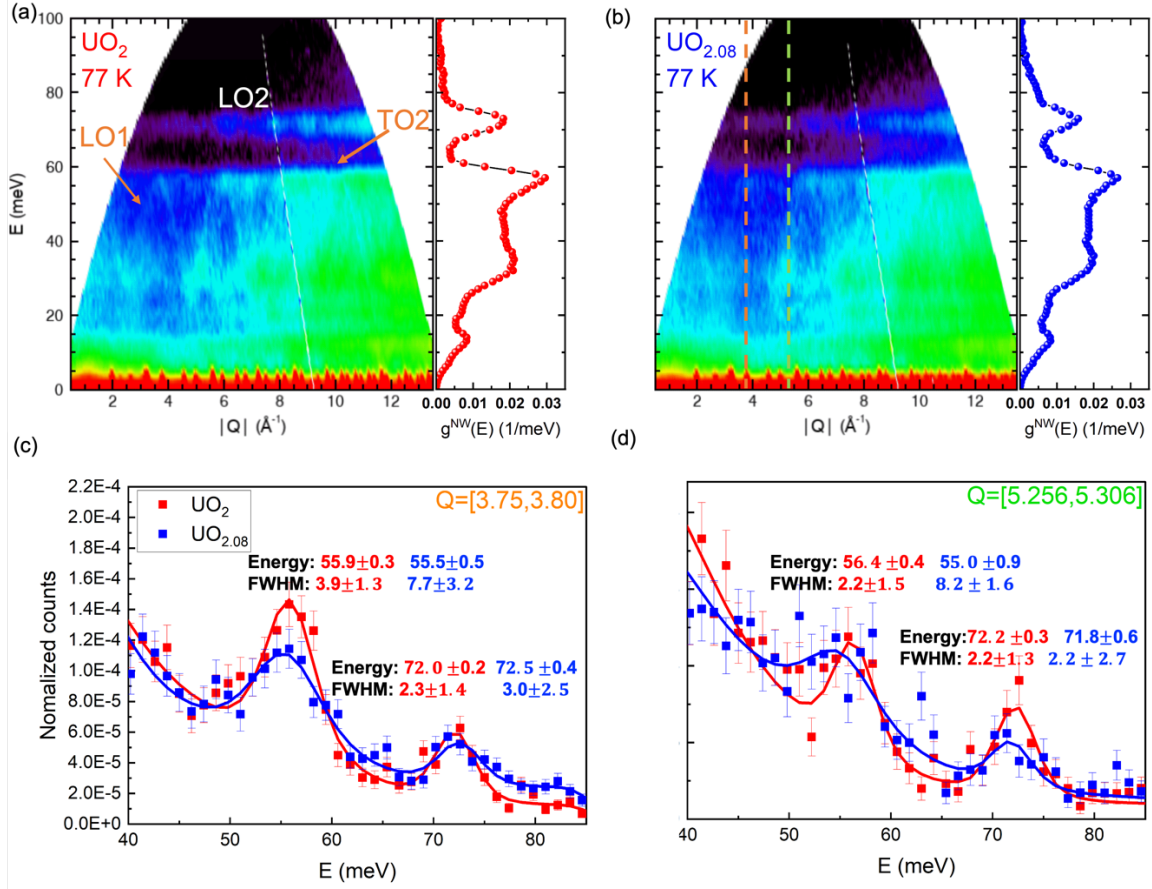

Fig. 3. Color contour plots of the scattering function $S(Q,E)$ as a function of E and Q and their corresponding neutron-weighted PDOS for (a) $UO_2$ and (b) $UO_{2.08}$ powders at 77K measured by INS. Scattering function as a function of energy at (c) Q=[3.75,3.8] and (d) Q=[5.256,5.306], corresponding to orange and blue dashed lines in the color contour plot (b). Red and blue lines in (c) and (d) are theoretical fittings of these INS data. Fitted phonon energies and linewidths are shown in the insert. Red and blue denote $UO_2$ and $UO_{2.08}$, respectively.

Thermal conductivity ($k$) of single crystals in the phonon gas model is quantitatively defined by
$$k = 1/3 C_V^{ph} V_g \tau^2 \quad (4)$$
where $C_V^{ph}$, $V_g$, and $\tau$ are the phonon heat capacity, the average phonon group velocity, and the average phonon lifetimes, respectively. With the similar $C_V^{ph}$ and $V_g$ in $UO_2$ and $UO_{2.08}$, the lower thermal conductivity in $UO_{2.08}$ must result from smaller average phonon lifetimes, i.e. higher phonon scattering rates than $UO_2$. For example, given that the thermal conductivity of $UO_{2.08}$ is 67% the value of $UO_2$ at 77 K, the average phonon lifetimes in $UO_{2.08}$ are 82% the value of $UO_2$. To validate this, we plot the measured scattering function $S(Q,E)$ as a function of $E$ and $Q$ at 77 K in Fig. 3a and 3b. ($S(Q,E)$ at 295 K can be found in Fig. S3 of SM). Interestingly, both $S(Q,E)$ contour plots exhibit clear sinusoidal shapes at the energy range of 20-55 meV and flat shapes at 56 meV and 72 meV, corresponding to LO1, TO2 and LO2 phonon branches along [100] direction. Moreover, LO1 phonon branches in $UO_{2.08}$ are broader than $UO_2$, which represents larger linewidths or higher phonon scattering rates in $UO_{2.08}$. Note that the LO1 phonon branches were found to contribute to the largest amount (>30%) of total thermal conductivity in $UO_2$.[17] Further energy cuts at Q=[3.75,3.8] and Q=[5.256,5.306] are shown in Fig. 3c and 3d. The fitted phonon energies of TO2 and LO2 modes in $UO_2$ are very close to $UO_{2.08}$, which reiterates the similar group velocities in them. The fitted phonon linewidths of TO2 and LO2 modes in $UO_2$ are 3.9±1.3 meV and 2.3±1.4 meV at Q=[3.75,3.8], and 2.2±1.5 meV and 2.2±2.7 meV at Q=[5.256,5.306], comparable to previously measured linewidths (2-3 meV) in $UO_2$ single crystal[17]. The phonon lifetimes (inverse of linewidths) of TO2 in $UO_{2.08}$ are 0.52 ps at Q=[3.75,3.8] and 0.50 ps at Q=[5.256,5.306], which are 49% of 1.06 ps and 27% of 1.88 ps in $UO_2$. In contrast, phonon lifetimes of LO2 in $UO_{2.08}$ are 1.80 ps at Q=[3.75,3.8] and 1.88 ps at Q=[5.256,5.306], which are almost the same as the corresponding values in $UO_2$ (1.38 ps at Q=[3.75,3.8] and 1.88 ps at Q=[5.256,5.306]). In brief, with similar $C_V^{ph}$ and $V_g$ in $UO_2$ and $UO_{2+x}$, small phonon lifetimes, i.e. high phonon scattering rates, lead to much lower thermal conductivity in $UO_{2+x}$.

## 4. Conclusion

In summary, we performed thermal conductivity and INS measurements on $UO_2$ and $UO_{2+x}$ at low temperatures (2-300 K). It is found that the thermal conductivities of all $UO_{2+x}$ single crystals exhibit a double-peak behavior where maxima occur at 10 and 220 K and a minimum occurs at the Néel temperature $T_N$= 30.8 K, consistent with previous studies on $UO_2$. Except in the vicinity of $T_N$, the thermal conductivity of $UO_{2+x}$ is significantly suppressed compared to $UO_2$ and decreases with the increasing of $x$. At $T_N$ = 30.8 K, the thermal conductivity of $UO_{2+x}$ is essentially independent of the $x$. Further INS measurements of the PDOS and scattering function demonstrate that heat capacity and phonon group velocities of $UO_2$ and $UO_{2+x}$ are similar and thus the suppressed thermal conductivity is attributed to high phonon scattering rates in $UO_{2+x}$. The fundamental new knowledge gained from this study may guide the future design of safe and efficient nuclear reactors.

**Acknowledgement**
H. M., M.S.B., K.G. and M.E.M. were supported by the Center for Thermal Energy Transport under Irradiation, an Energy Frontier Research Center funded by the U.S. Department of Energy (DOE), Office of Science, United States, Office of Basic Energy Sciences. D.J.A. and K.G. acknowledge support from Advanced Fuel Campaign program (INL). Portions of this research used resources at the Spallation Neutron Source, a U.S. DOE Office of Science User Facility operated by the Oak Ridge National Laboratory. We are grateful to David A. Andersson and Christopher R. Stanek for providing $UO_{2+x}$ crystals for the thermal conductivity measurements.

The data that supports the findings of this study are available within the article and its supplementary material.

**Reference**
[1] D.H. Hurley, A. El-Azab, M.S. Bryan, M.W.D. Cooper, C.A. Dennett, K. Gofryk, L. He, M. Khafizov, G.H. Lander, M.E. Manley, J.M. Mann, C.A. Marianetti, K. Rickert, F.A. Selim, M.R. Tonks, J.P. Wharry, Thermal Energy Transport in Oxide Nuclear Fuel, Chemical Reviews, DOI 10.1021/acs.chemrev.1c00262(2021).
[2] W.T. Thompson, B.J. Lewis, E.C. Corcoran, M.H. Kaye, S.J. White, F. Akbari, Z. He, R. Verrall, J.D. Higgs, D.M. Thompson, T.M. Besmann, S.C. Vogel, Thermodynamic treatment of uranium dioxide based nuclear fuel, International Journal of Materials Research, 98 (2007) 1004-1011.

[3] D.R. Olander, Mechanistic interpretations of UO2 oxidation, Journal of Nuclear Materials, 252 (1998) 121-130.
[4] J. Wang, R.C. Ewing, U. Becker, Average structure and local configuration of excess oxygen in UO2+x, Scientific Reports, 4 (2014) 4216.
[5] B.T.M. Willis, Positions of the Oxygen Atoms in UO2.13, Nature, 197 (1963) 755-756.
[6] B. Willis, The defect structure of hyper-stoichiometric uranium dioxide, Acta Crystallographica Section A: Crystal Physics, Diffraction, Theoretical and General Crystallography, 34 (1978) 88-90.
[7] D. Manara, C. Ronchi, M. Sheindlin, M. Lewis, M. Brykin, Melting of stoichiometric and hyperstoichiometric uranium dioxide, Journal of Nuclear Materials, 342 (2005) 148-163.
[8] F.J. Hetzler, E.L. Zebroski, THERMAL CONDUCTIVITY OF STOICHIOMETRIC AND HYPOSTOICHIOMETRIC URANIUM OXIDE AT HIGH TEMPERATURES, Transactions of the American Nuclear Society, DOI (1964).
[9] L.A. Goldsmith, J.A.M. Douglas, Measurements of the thermal conductivity of uranium dioxide at 670 1270 k, Journal of Nuclear Materials, 47 (1973) 31-42.
[10] M. Amaya, T. Kubo, Y. Korei, Thermal Conductivity Measurements on UO2+x from 300 to 1,400 K, Journal of Nuclear Science and Technology, 33 (1996) 636-640.
[11] J.T. White, A.T. Nelson, Thermal conductivity of UO2+x and U4O9−y, Journal of Nuclear Materials, 443 (2013) 342-350.
[12] S. Yamasaki, T. Arima, K. Idemitsu, Y. Inagaki, Evaluation of Thermal Conductivity of Hyperstoichiometric UO2+x by Molecular Dynamics Simulation, International Journal of Thermophysics, 28 (2007) 661-673.
[13] X.Y. Liu, M.W.D. Cooper, K.J. McClellan, J.C. Lashley, D.D. Byler, B.D.C. Bell, R.W. Grimes, C.R. Stanek, D.A. Andersson, Molecular Dynamics Simulation of Thermal Transport in ${\mathrm{UO}}_{2}$ Containing Uranium, Oxygen, and Fission-product Defects, Physical Review Applied, 6 (2016) 044015.
[14] D.L. Abernathy, M.B. Stone, M.J. Loguillo, M.S. Lucas, O. Delaire, X. Tang, J.Y.Y. Lin, B. Fultz, Design and operation of the wide angular-range chopper spectrometer ARCS at the Spallation Neutron Source, Review of Scientific Instruments, 83 (2012) 015114.
[15] J.W.L. Pang, A. Chernatynskiy, B.C. Larson, W.J.L. Buyers, D.L. Abernathy, K.J. McClellan, S.R. Phillpot, Phonon density of states and anharmonicity of ${\mathrm{UO}}_{2}$, Physical Review B, 89 (2014) 115132.
[16] M.S. Bryan, J.W.L. Pang, B.C. Larson, A. Chernatynskiy, D.L. Abernathy, K. Gofryk, M.E. Manley, Impact of anharmonicity on the vibrational entropy and specific heat of ${\mathrm{UO}}_{2}$, Physical Review Materials, 3 (2019) 065405.
[17] J.W.L. Pang, W.J.L. Buyers, A. Chernatynskiy, M.D. Lumsden, B.C. Larson, S.R. Phillpot, Phonon Lifetime Investigation of Anharmonicity and Thermal Conductivity of ${\mathrm{UO}}_{2}$ by Neutron Scattering and Theory, Physical Review Letters, 110 (2013) 157401.
[18] T. Arima, S. Yamasaki, Y. Inagaki, K. Idemitsu, Evaluation of thermal properties of UO2 and PuO2 by equilibrium molecular dynamics simulations from 300 to 2000K, Journal of Alloys and Compounds, 400 (2005) 43-50.
[19] P. Goel, N. Choudhury, S.L. Chaplot, Fast ion diffusion, superionic conductivity and phase transitions of the nuclear materials UO2 and Li2O, Journal of Physics: Condensed Matter, 19 (2007) 386239.
[20] D.C. Wallace, Statistical Physics of Crystals and Liquids.

[21] K. Gofryk, S. Du, C.R. Stanek, J.C. Lashley, X.Y. Liu, R.K. Schulze, J.L. Smith, D.J. Safarik, D.D. Byler, K.J. McClellan, B.P. Uberuaga, B.L. Scott, D.A. Andersson, Anisotropic thermal conductivity in uranium dioxide, Nature Communications, 5 (2014) 4551.
[22] J.P. MOORE, D.L. MCELROY, Thermal Conductivity of Nearly Stoichiometric Single-Crystal and Polycrystalline UO2, Journal of the American Ceramic Society, 54 (1971) 40-46.
[23] K. Shrestha, T. Yao, J. Lian, D. Antonio, M. Sessim, M.R. Tonks, K. Gofryk, The grain-size effect on thermal conductivity of uranium dioxide, Journal of Applied Physics, 126 (2019) 125116.
[24] J.D. Higgs, W.T. Thompson, B.J. Lewis, S.C. Vogel, Kinetics of precipitation of U4O9 from hyperstoichiometric UO2+x, Journal of Nuclear Materials, 366 (2007) 297-305.
[25] R.I. Palomares, M.T. McDonnell, L. Yang, T. Yao, J.E.S. Szymanowski, J. Neuefeind, G.E. Sigmon, J. Lian, M.G. Tucker, B.D. Wirth, M. Lang, Oxygen point defect accumulation in single-phase $\mathrm{U}{\mathrm{O}}_{2+x}$, Physical Review Materials, 3 (2019) 053611.
[26] M.S. Bryan, L. Fu, K. Rickert, D. Turner, T.A. Prusnick, J.M. Mann, D.L. Abernathy, C.A. Marianetti, M.E. Manley, Nonlinear propagating modes beyond the phonons in fluorite-structured crystals, Communications Physics, 3 (2020) 217.
[27] J.J. Huntzicker, E.F. Westrum, The magnetic transition, heat capacity, and thermodynamic properties of uranium dioxide from 5 to 350 K, The Journal of Chemical Thermodynamics, 3 (1971) 61-76.
[28] G.D. Khattak, Specific heat of uranium dioxide (UO2) between 0.3 and 50 K, physica status solidi (a), 75 (1983) 317-321.
[29] R. De Batist, R. Gevers, M. Verschueren, Magnon Contribution to the Low-Temperature Specific Heat of UO2, physica status solidi (b), 19 (1967) 77-88.
[30] A.P. Cracknell, S.J. Joshua, The Spin-Wave Dispersion Relations and the Spin-Wave Contribution to the Specific Heat of Antiferromagnetic UO2, physica status solidi (b), 36 (1969) 737-745.

Supplementary Materials

# High Phonon Scattering Rates Suppress Thermal Conductivity in Hyperstoichiometric Uranium Dioxide

Hao Ma,[1*] Matthew S. Bryan,[1] Judy W. L. Pang,[1] Douglas L. Abernathy,[2] Daniel J. Antonio,[3] Krzysztof Gofryk,[3] and Michael E. Manley[1*]

[1]*Materials Science and Technology Division, Oak Ridge National Laboratory, Oak Ridge, Tennessee 37831, USA*
[2]*Neutron Scattering Division, Oak Ridge National Laboratory, Oak Ridge, Tennessee 37831, USA*
[3]*Idaho National Laboratory, Idaho Falls, Idaho 83415, USA*

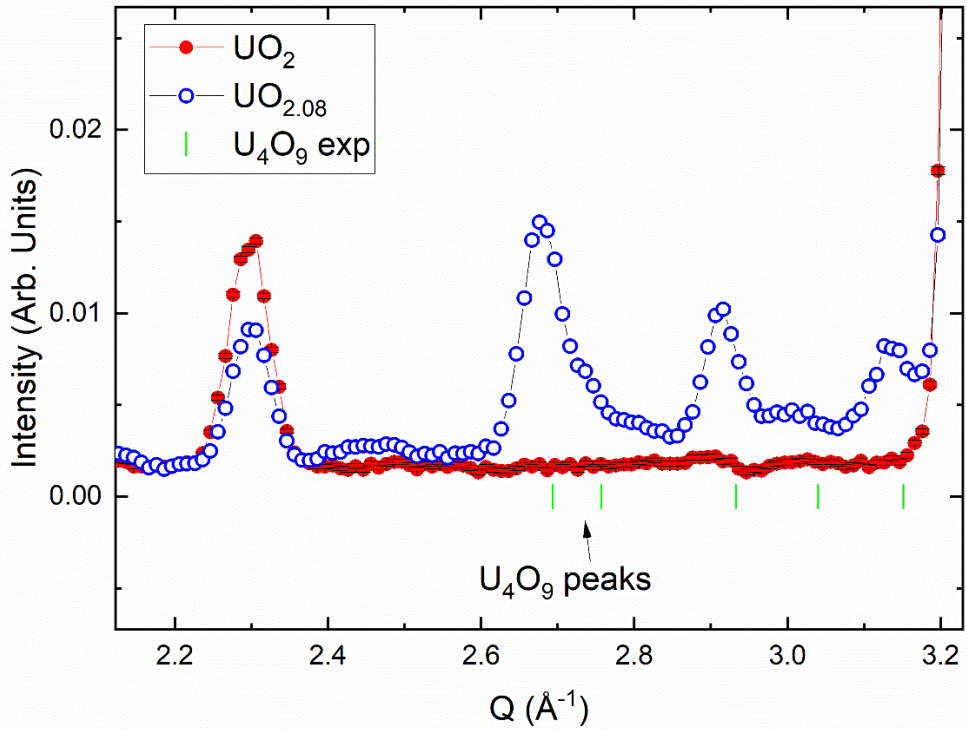

**Fig. S1.** The measured neutron diffraction patterns of $UO_2$ and $UO_{2.08}$ at 295 K. $U_4O_9$ data is from Ref. 1.

* Corresponding author. Email: mah1@ornl.edu
* Corresponding author. Email: manleyme@ornl.gov

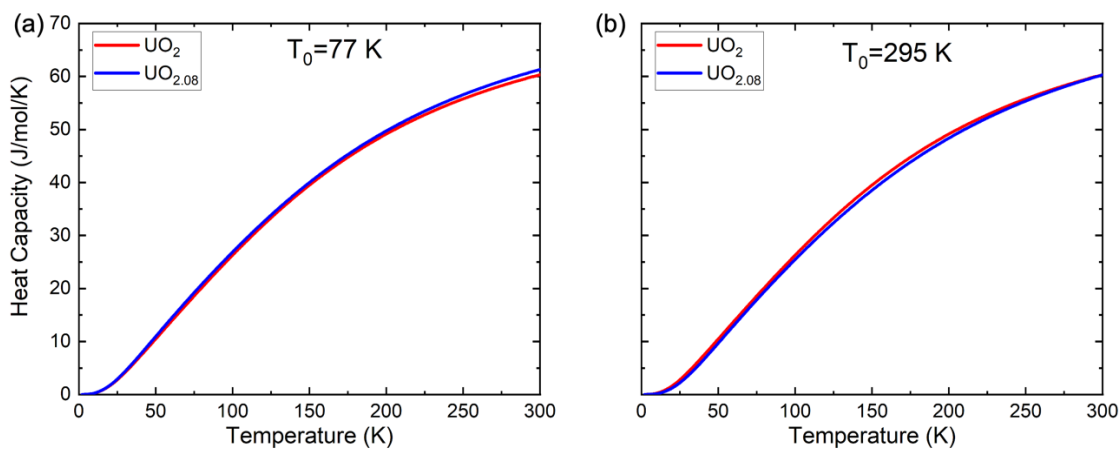

**Fig. S2.** The heat capacity ($C_V^{ph}$) of UO$_2$ (red line) and UO$_{2.08}$ (blue line) with respect to temperature were calculated using unweighted PDOS at (c) T$_0$=77 K and (d) T$_0$=295 K.

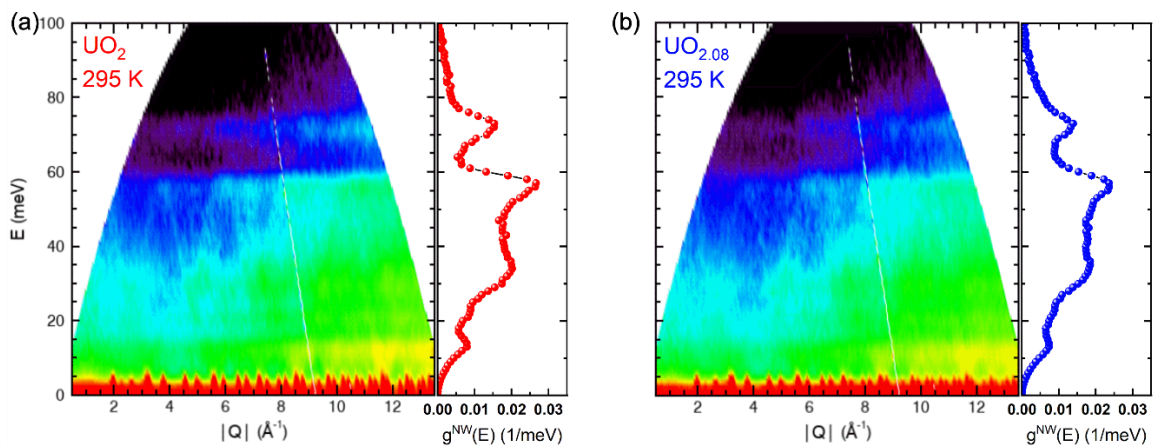

**Fig. S3.** Color contour plots of the scattering function S(Q, E) as a function of E and Q and their corresponding neutron-weighted PDOS for (a) UO$_2$ and (b) UO$_{2.08}$ at 295 K measured by INS.

**References**

1. Garrido, F*., et al.*, *Inorganic Chemistry* (2006) **45** (20), 8408